\documentclass[twocolumn,english,amsart,showpacs,,showkeys,preprintnumbers,amsmath,amssymb,floatfix]{revtex4-1}

\usepackage{tikz,xcolor}
\usepackage[colorlinks = true,
linkcolor = blue,
urlcolor  = blue,
citecolor = blue,
anchorcolor = blue]{hyperref}
\definecolor{lime}{HTML}{A6CE39}
\DeclareRobustCommand{\orcidicon}{%
	\begin{tikzpicture}
		\draw[lime, fill=lime] (0,0)
		circle [radius=0.16]
		node[white] {{\fontfamily{qag}\selectfont \tiny ID}};
		\draw[white, fill=white] (-0.0625,0.095)
		circle [radius=0.007];
	\end{tikzpicture}
	\hspace{-2mm}
}
\foreach \x in {A, ..., Z}{%
	\expandafter\xdef\csname orcid\x\endcsname{\noexpand\href{https://orcid.org/\csname orcidauthor\x\endcsname}{\noexpand\orcidicon}}
}

\usepackage[T1]{fontenc}
\usepackage[latin9]{inputenc}
\usepackage{color}
\usepackage{array}
\usepackage{amstext}
\usepackage{graphicx}
\usepackage{esint}
\usepackage{rotating}
\usepackage[font=small,labelfont=bf]{caption}
\usepackage{xcolor}

\makeatletter

\@ifundefined{textcolor}{}
{%
 \definecolor{BLACK}{gray}{0}
 \definecolor{WHITE}{gray}{1}
 \definecolor{RED}{rgb}{1,0,0}
 \definecolor{GREEN}{rgb}{0,1,0}
 \definecolor{BLUE}{rgb}{0,0,1}
 \definecolor{CYAN}{cmyk}{1,0,0,0}
 \definecolor{MAGENTA}{cmyk}{0,1,0,0}
 \definecolor{YELLOW}{cmyk}{0,0,1,0}
 }

\@ifundefined{definecolor}
 {\usepackage{color}}{}
\@ifundefined{definecolor}
 {\usepackage{color}}{}
\makeatother
\usepackage{babel}

\begin{document}


\title {Self-Similar Properties of the Proton Structure at Low x  within the \textbf{xFitter} framework  }

\author {Shahin Atashbar Tehrani$^{1,2}$\orcidB{}}
\email{Atashbar@ipm.ir}

\author {Fatemeh Taghavi-Shahri$^{3}$\orcidA{}}
\email{taghavishahri@um.ac.ir}

\author {Samira Shoeibi Mohsenabadi$^{3}$\orcidC{}}
\email{Samira.Shoeibi@ipm.ir}

\affiliation {
$^{(1)}$School of Particles and Accelerators, Institute for Research in Fundamental Sciences (IPM), P.O.Box 19395-5531, Tehran, Iran.\\
$^{(2)}$Department of Physics, Faculty of Nano and Bio Science and Technology, Persian Gulf University, 75169 Bushehr, Iran. \\
$^{(3)}$Department of Physics, Ferdowsi University of Mashhad, P.O.Box 1436, Mashhad, Iran
}

\date{\today}

%
\begin{abstract}\label{abstract}

{ 
The structure of the proton exhibits  Fractal behavior at low  \textit{x}, where \textit{x} is the fraction of the proton's momentum carried by the interacting partons. This Fractal behavior is characterized by self-similar properties at different scales and can be quantified using the concept of Fractal dimension. An investigation into the Fractal properties of the proton structure at low \textit{x}  is critical for understanding the fundamental properties of the strong force and developing a more comprehensive understanding of the hadron structure. Fractals, characterized by self-similar patterns across scales, demonstrate a direct correlation between their Fractal dimension and entropy, where higher Fractal dimensions correspond to increased informational content.  Furthermore, it is essential for designing high-energy physics experiments and developing more accurate models of subatomic particle interactions. This paper has a fresh look at the self-similar properties of the proton structure at low \textit{x}. Our study involves the use of the \textbf{xFitter} framework to parameterize the proton structure functions with a Fractal formalism at low \textit{x}. We also examine how the inclusion of new data affects the results of our analysis.

}

\end{abstract}
\maketitle

%
%
\section{Introduction}\label{sec:sec1}
The proton is a subatomic particle that is composed of smaller particles called quarks and gluons. At high energies, the structure of the proton can be described using the theory of quantum chromodynamics (QCD), which predicts that a strong force binds together the quarks and gluons.\\
At low energies or low momentum transfers, the proton structure becomes more complex and difficult to describe only using QCD. 
The proton structure function, $F_2(x, Q^2)$, goes up with $Q^2$ at low Bjorken \textit{x} (the fraction of the proton's momentum carried by the interacting parton), which indicates rapid growth of the partons at low \textit{x}. The experiments show that the number of  partons inside a proton goes up at low $\textit{x}$, and falls at high $\textit{x}$ \cite{H1:2001ert,Abramowicz:2015mha,H1:2010fzx,Aaron:2009aa}. 
Proton has three valence quarks that are more significant at low $Q^2$, but at high $Q^2$, the sea quarks, which are quark-antiquark pairs, increase in number and have an important role at small  \textit{x}. Because the knowledge of these densities at a much smaller value of $\textit{x}$ will be needed for any collider predictions, then  it is crucial to know how PDFs behave at low $\textit{x}$.\\
It is shown that  the structure of the proton can be  characterized by the concept of Fractals, which are objects that exhibit self-similar properties at different scales \cite{Lastovicka:2002hw, Lastovicka:2004mq, Choudhury:2003yy, Jahan:2014sqa, Choudhury:2016fjy,  Mohsenabadi:2021vvj}. Fractals possess a holographic nature too. Intriguingly, Fractals are members of a fundamental class of statistical objects, with their basis lying at the heart of information theory.
The holographic principle is a concept in theoretical physics that suggests that the information content of a certain region of space can be fully encoded on its boundary.  Fractals manifest in various aspects of nature. Scientists were led by Fractals to the realization that apparent chaos in fact structured. It appears that the complex phenomena have a hidden order. The holographic principle and Fractals exhibit an intricate interplay, underpinned by their shared understanding of information organization in the universe. Fractals, characterized by self-similar patterns across scales, demonstrate a direct correlation between their Fractal dimension and entropy, where higher Fractal dimensions correspond to increased informational content. This connection is rooted in information theory, which posits that complex, self-similar systems, like Fractals, can be more efficiently encoded. The holographic principle builds upon this, proposing that the information within a given volume can be fully described by the information encoded on its boundary, akin to a hologram. Crucially, the Shannon entropy of a system, a fundamental information-theoretic measure, is directly related to its Fractal dimension, suggesting the universe's underlying structure may be Fractal-like, enabling the efficient storage and transmission of information as envisaged by the holographic principle \cite{Mureika:2006tz}.\\
One way to quantify the Fractal properties of the proton structure at low \textit{x} is to use the concept of Fractal dimension in the PDFs parameterized by the Fractal distributions. The Fractal distributions are a type of probability distribution that is statistically self-similar and  look the same at different scales.
Understanding the Fractal patterns of the proton at low \textit{x} is essential for learning more about the basic properties of the strong force and the hadron structure. It also affects high-energy physics experiments that rely on the proton structure to predict the results of particle collisions.\\
In this  paper, we have  a fresh look at the self-similar properties of the proton structure at low \textit{x}. We used the \textbf{xFitter} framework~\cite{Sapronov:2015pba} to determine the Fractal PDFs and proton's reduced cross-section at the low \textit{x} region by including a selection of data sets (500 data points) from the combined HERA I+II deep-inelastic scattering data~\cite{Abramowicz:2015mha}. The various PDF fits are performed to four different data sets: (i) inclusive HERA I+II cross-section data at 920 (GeV), (ii) inclusive HERA I+II cross-section data at 920 and 820 (GeV), (iii) inclusive HERA I+II cross-section data  at 920, 820, and 575 (GeV) (iv) inclusive HERA I+II cross-section data  at 920, 820, 575 and 460 (GeV). Then, the impact of adding each data set on the proton's reduced cross-section is investigated.\\
This paper is organized as follows: In Section 2, we provide a concise overview of the Fractal dimension and the Fractal distributions that we need to study  the proton structure. Section \ref{parametrization} describes the parametrization of the structure-function based on Fractal formalism, which is a key quantity used in this analysis. Next, in Section \ref{fitting}, we discuss the fitting procedures used in the QCD analysis of the data. Finally, in Section \ref{summary}, we summarize the key findings of the study and present our conclusions regarding the implications of our results.
Following this structure, we aim to provide readers with a clear and comprehensive understanding of the theoretical concepts, data analysis techniques, and results presented in this paper.

\section{The Fractal Dimension  and the Fractal Distribution}
\label{dimension}
The concept of the Fractal dimension requires an understanding of the meaning of the term "dimension." In non-fractional dimensions, the number of dimensions corresponds to the number of independent directions in a corresponding coordinate system. For example, as shown in Fig. \ref{dimension}  a line has  dimension 1, a square has  dimension 2, and a cube has  dimension 3. However, Fractal objects such as the Sierpinski gasket, shown in Fig. \ref{gasket}, require a more general definition of dimension than traditional Euclidean geometry. 
It is shown in Fig. \ref{dimension} that the  traditional notions of geometry for defining dimension can be linked to the self-similarity property by this relation:
\begin{equation}\label{Eq:Fractal-Dimention1}
	D = \frac{log(number~of~self~ similar~objects)}{log(magnification~factor)} = \frac{log(N)}{log(r)}
\end{equation}
here \textit{r} is the number of each side segment (Magnification factor) and \textit{N} is the number of self-similar shapes created by the segmentation. It means when a line is divided in the middle, two lines of half-length are obtained. By magnifying one of them by a factor of two, the original line can be reconstructed. The same principle applies to dividing a square into four smaller squares or a cube into smaller cubes. Similarly, as another example, if the magnification factor for a square is 3, then the number of smaller squares will be $3^2=9$. In a similar fashion, a cube magnified by a factor of 3 will result in $3^3=27$ smaller cubes. According to this relation, for example for 3 segmentation case, we would obtain the values of D=$\frac{log (3)}{log (3)}$=1 for the dimension of line, D=$\frac{log (3^{2})}{log (3)}$=2 and D=$\frac{log (3^{3})}{log (3)}$=3 for the dimensions of square and cube, respectively. \\
To quantify the dimension of a Fractal object, one cannot simply use the definition existing for dimension, as this approach fails to capture the complex and self-similar structure of Fractals. Instead, the concept of the Fractal dimension provides a more nuanced definition that captures the self-similarity and irregularity of Fractal objects. The Fractal dimension of an object can be thought of as a measure of how much space it fills, relative to its size.
The Sierpinski gasket, for example, is a Fractal object that is formed by repeatedly removing triangles from a larger triangle. As the process is repeated infinitely many times, the Sierpinski gasket exhibits self-similarity at different scales. Using Eq. (\ref{Eq:Fractal-Dimention1}), the Fractal dimension of the Sierpinski gasket is approximately 1.585, which is a non-integer value that reflects its complex and irregular structure. In iteration No. 1, for example, we have
\begin{equation}
	{\cal D}=\frac{\log 3}{\log 2 }\approx 1.585.
	\label{dimdef3}
\end{equation}
Fractal objects, on the other hand, have non-integer dimensions and exhibit self-similarity at all scales. In contrast to traditional geometric shapes, real Fractals in nature do not exhibit the same degree of self-similarity for all magnification factors, and the number of self-similar objects changes in a more complex manner. The concept of the Fractal dimension provides a way to quantify the self-similarity and irregularity of the Fractal objects.
\begin{figure}[!htb]
	\begin{center}
		\includegraphics[clip,width=0.45\textwidth]{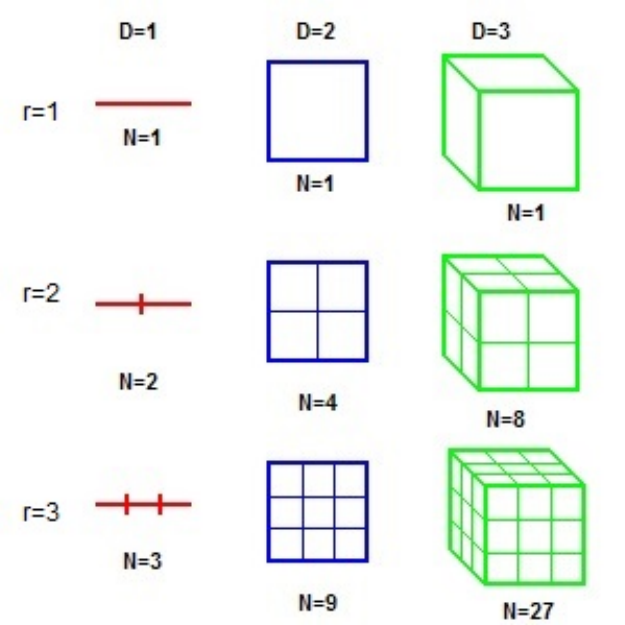}
		\vspace{0.0cm}
		\caption{{\small Regular self-similarity in Eclidian space} 
			\label{dimension}}
	\end{center}
\end{figure}

\begin{figure}[!htb]
	\begin{center}
		\vspace{-10.0cm}
		\includegraphics[clip,width=0.55\textwidth]{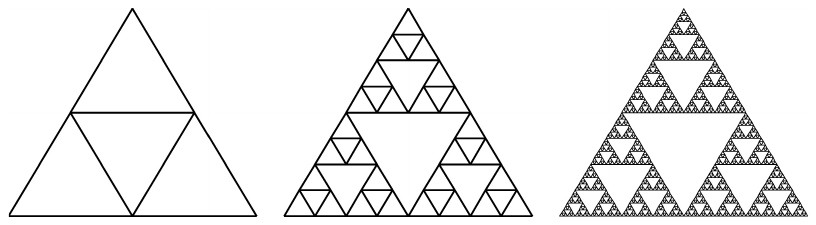}
		\vspace{-1.0cm}
		\caption{{\small Sierpinski gasket Fractal in iterations No. 1, 3 and 6 
				(from left). Iteration No. 1 corresponds to the {\it seed image} which is arbitrary while the iteration always converges to the same object.} 
			\label{gasket}}
	\end{center}
\end{figure}

\begin{figure*}[!htb]
	\begin{center}
		\includegraphics[clip,width=0.4\textwidth]{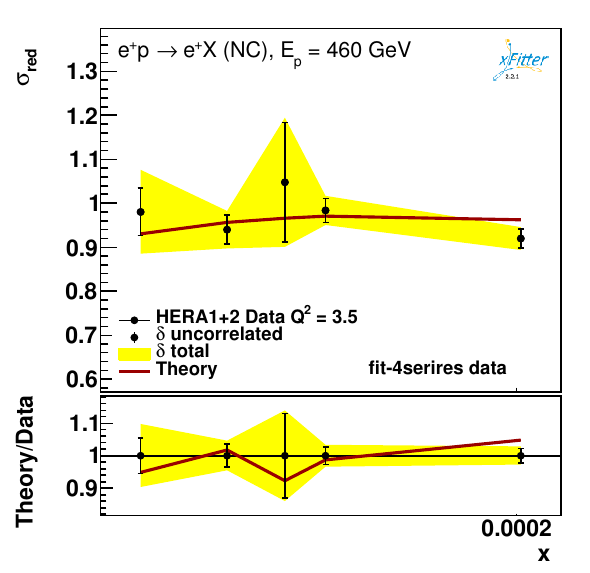 }
		\includegraphics[clip,width=0.4\textwidth]{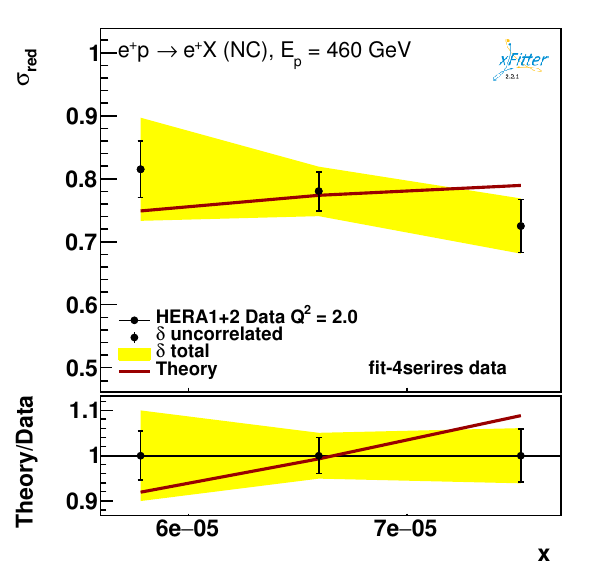 }
		\includegraphics[clip,width=0.4\textwidth]{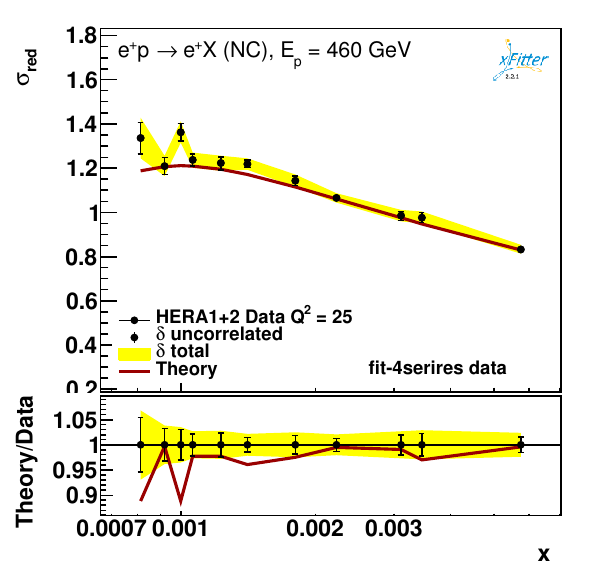 }
		\includegraphics[clip,width=0.4\textwidth]{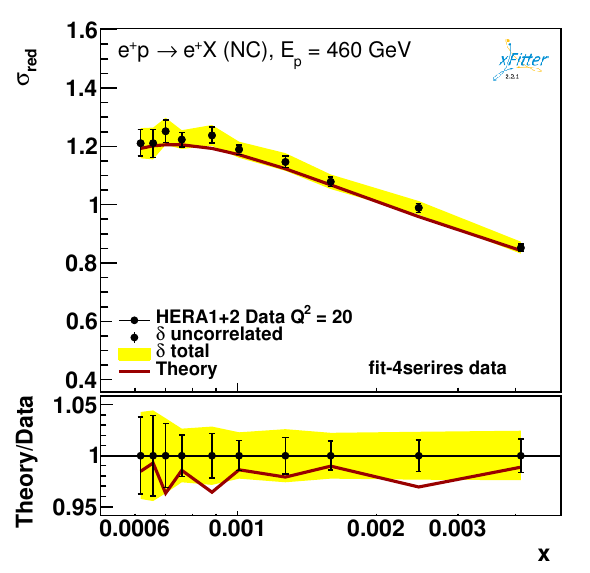 }
		\caption{{\small  The prediction for the reduced cross-section in the low \textit{x} region below $x<0.001$ for $Q^2=3.5 GeV^2$, $Q^2=2 GeV^2$ , $Q^2=20 GeV^2$ and  $Q^2=25 GeV^2$ ($E_p=460GeV$). Data points are from NC interactions in HERA  positron- proton DIS processes for fit-4serires data.} \label{fig:fig1}}
	\end{center}
\end{figure*}

\begin{figure*}[!htb]
	\begin{center}
		\includegraphics[clip,width=0.4\textwidth]{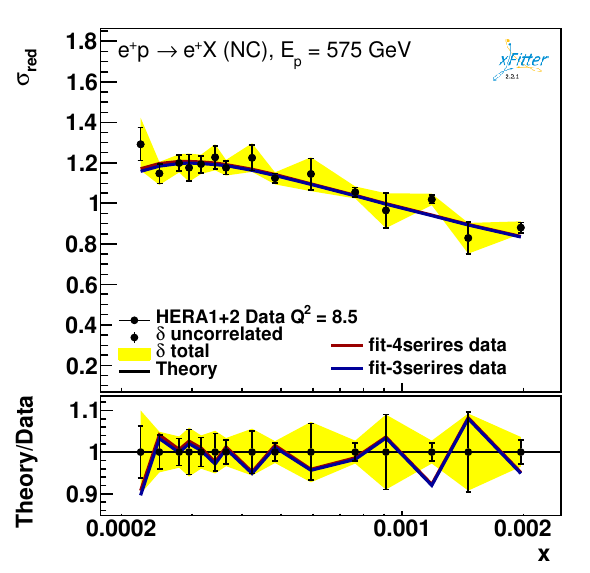 }
		\includegraphics[clip,width=0.4\textwidth]{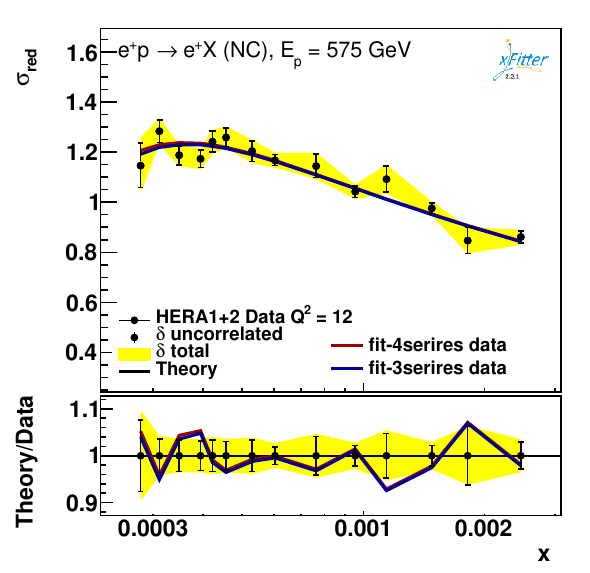 }
		\includegraphics[clip,width=0.4\textwidth]{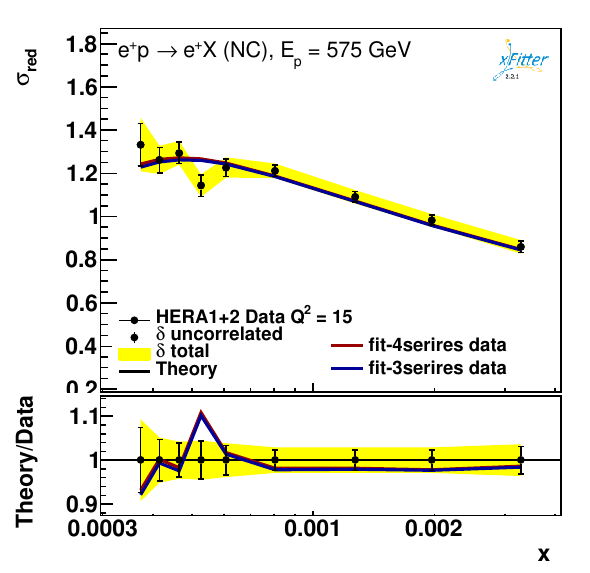 }
		\includegraphics[clip,width=0.4\textwidth]{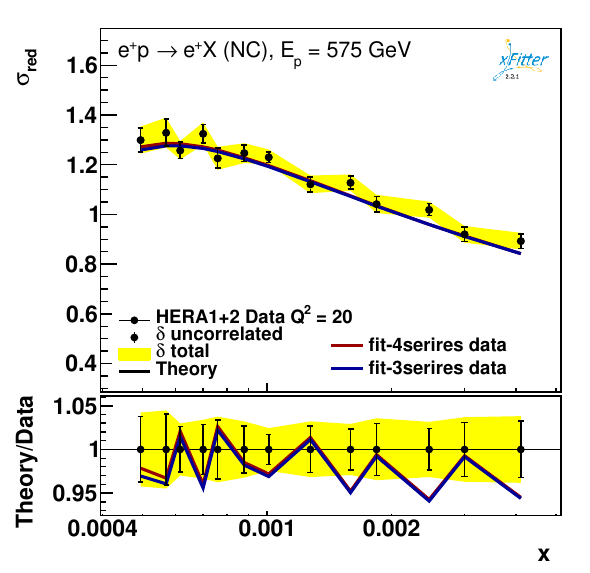 }
		\caption{{\small . The prediction for the reduced cross-section in the low \textit{x} region below $x<0.001$ for $Q^2=8.5 GeV^2$, $Q^2=12 GeV^2$, $Q^2=15 GeV^2$ and $Q^2=20 GeV^2$  ($E_p=575GeV$). Data points are from NC interactions in HERA  positron- proton DIS processes for fit-1serires data and fit-2serires data .} \label{fig:fig2}}
	\end{center}
\end{figure*}

\begin{figure*}[!htb]
	\begin{center}
		\includegraphics[clip,width=0.4\textwidth]{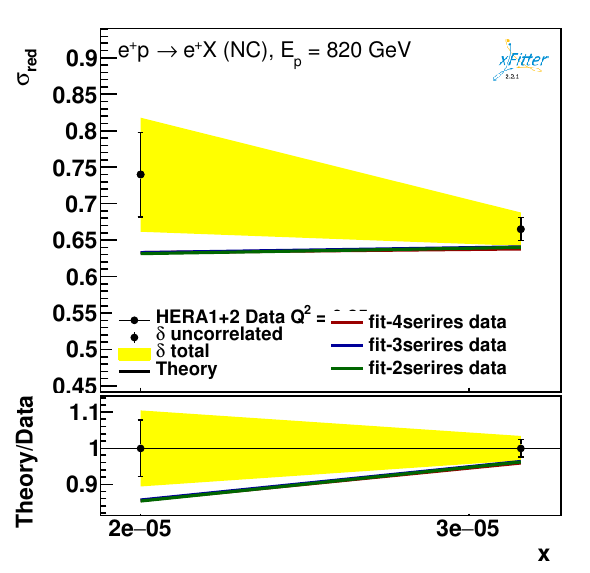 }
		\includegraphics[clip,width=0.4\textwidth]{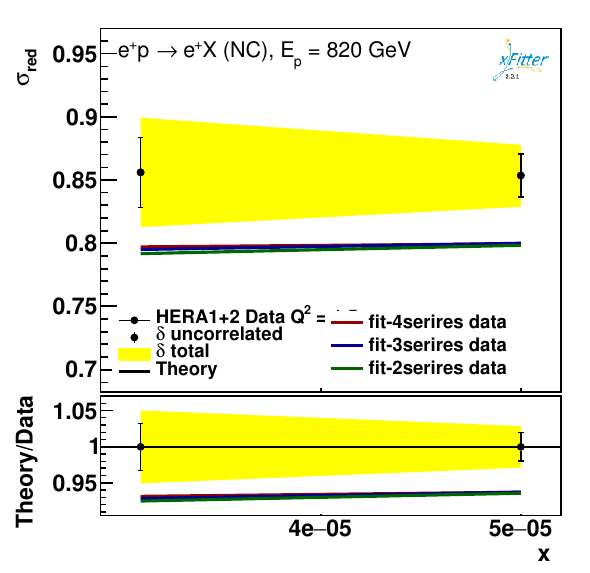 }
		\includegraphics[clip,width=0.4\textwidth]{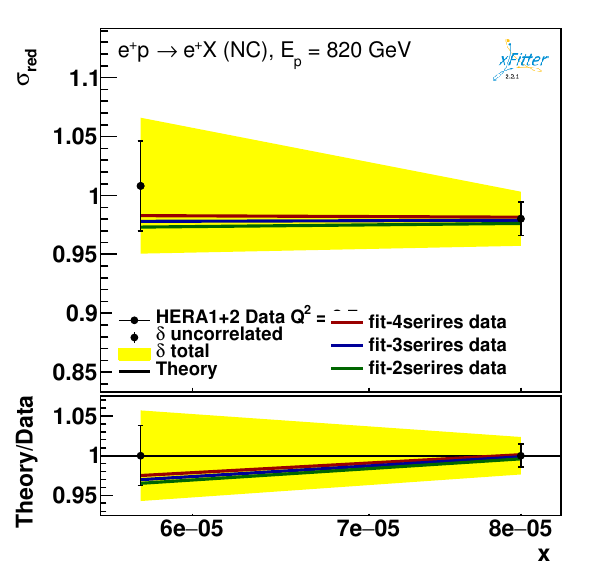 }
		\includegraphics[clip,width=0.4\textwidth]{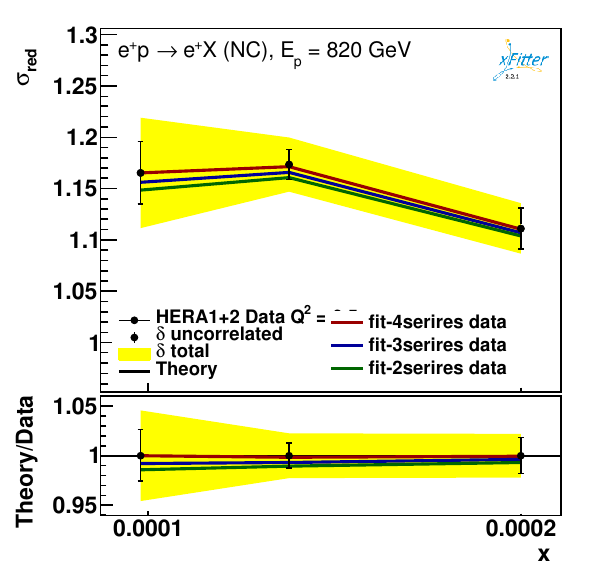 }
		\caption{{\small . The prediction for the reduced cross-section in the low \textit{x} region below $x<0.0001$ for  $E_p=820GeV$. Data points are from NC interactions in HERA  positron-proton DIS processes for fit-1serires data, fit-2serires data, and fit-3serires data.} \label{fig:fig3}}
	\end{center}
\end{figure*}

\begin{figure*}[!htb]
	\begin{center}
		\includegraphics[clip,width=0.4\textwidth]{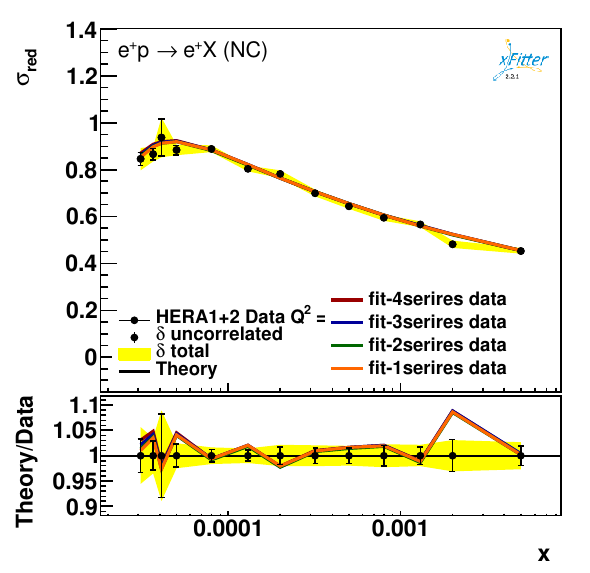 }
		\includegraphics[clip,width=0.4\textwidth]{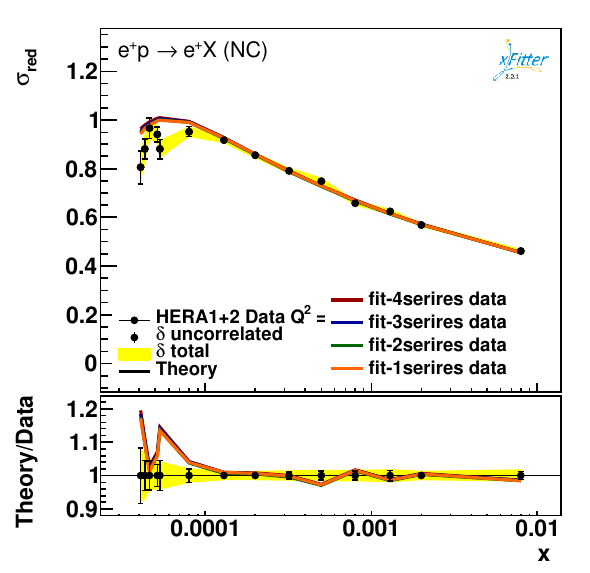 }
		\includegraphics[clip,width=0.4\textwidth]{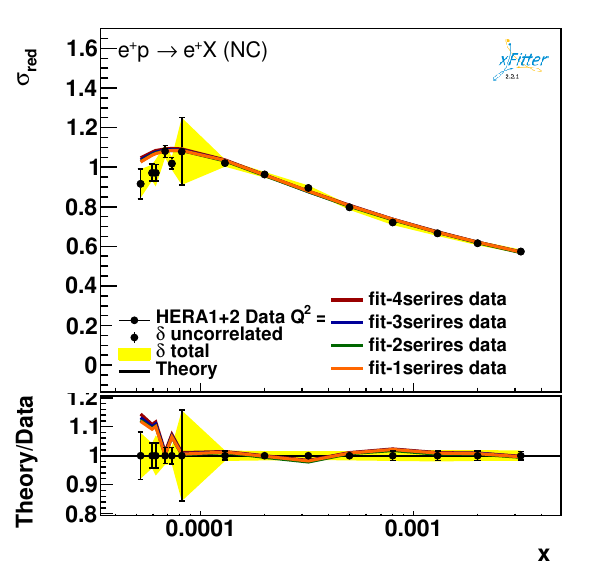 }
	   	\includegraphics[clip,width=0.4\textwidth]{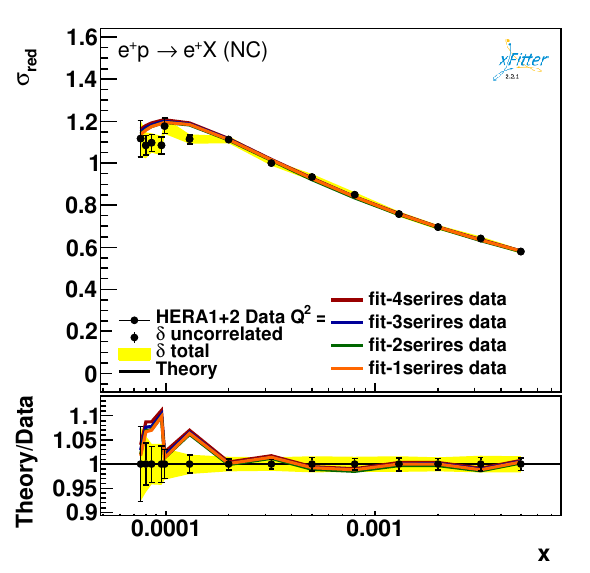 }
		\caption{{\small . The prediction for the reduced cross-section in the low \textit{x} region below $x<0.01$. Data points are from NC interactions in HERA  positron-proton DIS processes for fit-1serires data, fit-2serires data, fit-3serires data, and fit-4serires data.}  \label{fig:fig4}}
	\end{center}
\end{figure*}

\begin{figure}[!htb]
	\begin{center}
		\includegraphics[clip,width=0.45\textwidth]{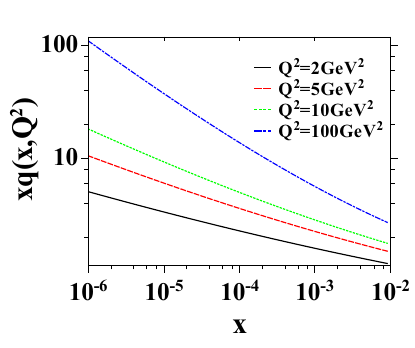 }
		\caption{{\small The $xq(x,Q^2)$ for $Q^2=2GeV^2$, $Q^2=5GeV^2$ , $Q^2=10GeV^2$ and $Q^2=100GeV^2$  for fit-4serires data .} 
			\label{xq}}
	\end{center}
\end{figure}

\begin{figure}[!htb]
	\begin{center}
		\includegraphics[clip,width=0.45\textwidth]{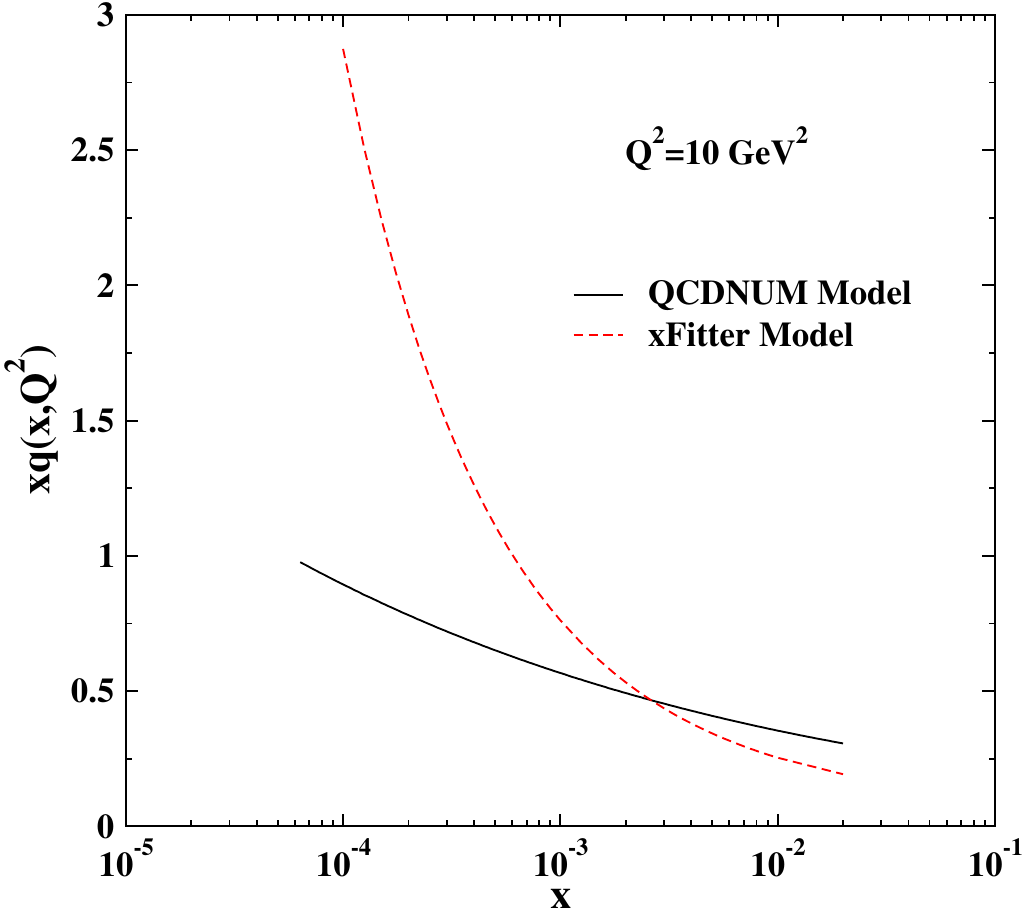 }
		\caption{{\small The $xq(x,Q^2)$  form Fractal model in xFitter (Lo approximation and only qurk PDfs) and those from  ~\cite{Mohsenabadi:2021vvj} (NLO approximation with quark and gluon PDFs).} 
			\label{xq2}}
	\end{center}
\end{figure}
The concept of the Fractal dimension can also be extended to non-discrete Fractals, where the magnification factor is a real number $z$ and the number of self-similar objects is represented by a density function $f(z)$. In this case, the dimension may vary with the scaling factor and a local dimension can be defined as the logarithmic derivative of the density function:
\begin{equation}\label{local-D}
	{\cal D}(z)=\lim_{z \rightarrow 0}\frac{\log f(z)}{\log
		z }
\end{equation}
Here, $f(z)$ represents the density of self-similar objects at scale $z$. We can rewrite Eq.~\ref{local-D} as:
\begin{equation}
	\log f(z)={\cal D} \cdot \log z + {\cal D}_0
	\label{powerlaw}
\end{equation}
where ${\cal D}_0$ defines the normalisation of $f(z)$, which thus has a {\it power law} behaviour, \mbox{$f(z) \propto z^D$.}\\
In general, Fractals may have two {\it independent} magnification factors, 
$z$ and $y$. In this case the density $f(z,y)$ is written in the following way~\cite{Lastovicka:2002hw, Lastovicka:2004mq}
\begin{equation}
	\log f(z,y)={\cal D}_{zy} \cdot \log z \cdot \log y + {\cal D}_z \cdot \log z + {\cal D}_y \cdot \log y + {\cal D}_0 .
	\label{twod}
\end{equation}
where $D_{zy}$ demonstrates the dimensional correlation related to two magnification factors, $y$ and $z$.\\
It is important to mention that there is a certain freedom in selecting
magnification factors without changing the shape of the function $f(z,y)$.
It is possible to use any non-zero power of a factor multiplied by a constant:
$z \rightarrow a z^\lambda$ , without affecting the underlying self-similarity
of the Fractal distributions. The only effect of such a change is a redefinition of the
dimensional parameters ${\cal D}_{{z,y,zy}}$ and of the normalization
${\cal D}_0$, respectively.

\begingroup
\begin{table*}[htp]
	\caption{\label{AllData} The list of all data sets: DIS HERA I+II used in the present analysis. For each data set, we 
		indicate process, measurement, reference and the ranges of their kinematic cuts such as $x$, $Q^2$ (GeV$^2$). }
	\begin{center}
		\begin{tabular}{l|l|c|p{3.39cm}p{3.39cm}}
			~Data set   & ~Experiment   & Ref.  & ~Kinematic ranges and details\\
			\hline 		
			\multicolumn{1}{l}{\bf{HERA I+II}}\\\hline	
			\hline 			
			$\sigma_{r,NC}$	 & ~HERA I+II NC $e^-p$ 460~ &\cite{Abramowicz:2015mha} &~$1\leq Q^2 \leq 100$ &$1\times 10^{-6}\leq x \leq$0.01\\ 
			&  ~HERA I+II NC $e^-p$ 575~ &\cite{Abramowicz:2015mha}&~$1\leq Q^2 \leq100$, &$1\times 10^{-6}\leq x \leq$0.01\\ 
			&  ~HERA I+II NC $e^+p$ 820~ &\cite{Abramowicz:2015mha}&~$1\leq Q^2 \leq100$, &$1\times 10^{-6}\leq x \leq$0.01\\ 
			&  ~HERA I+II NC $e^+p$ 920~ &\cite{Abramowicz:2015mha}&~$1\leq Q^2 \leq 100$, &$1\times 10^{-6}\leq x \leq$0.01 \\
			\hline			
		\end{tabular}
\end{center}
\end{table*}
\endgroup
It should be noted that the Fractal behavior is seen on the shape of the quark distribution functions or PDFs( parton distribution functions) at low Bjorken- x values, not in the shape of the proton.
The parton distribution functions at low x has a linear behavior in log-log space and a power-law behavior in ordinary phase space of  x and $Q^2$.  Then, we can describe  these PDFs with  Fractal distributions. In this work, we want to show that the quark distribution functions have a monofractal behavior at low x (and $Q^2$) which  means it has self-similar properties with fixed exponents (Fixed Fractal dimension). In the study of proton structure at small-x values, researchers have observed an increased rate of interactions between gluons, leading to enhanced densities of both gluons and sea quarks. This phenomenon exhibits fractal characteristics, as discussed here, which has sparked interest in applying fractal formalism to Parton Distribution Functions (PDFs).  The key concept behind this application lies in the recognition of scaling behaviors following a power law. A notable reference \cite{Abdulov:2021ivr} demonstrates that, when plotted on a log-log scale, the density functions of sea quarks, which depend on momentum transfer $Q^2$ and the Bjorken variable x specifically for $x < 0.01$ (outside the valence quark region), exhibit a linear trend. Consequently, the kinematical variables x and $Q^2$ emerge as suitable magnification factors. It is important to note that these magnification factors must be positive, non-zero, and dimensionless. Thus, in addition to x and $Q^2$, Reference \cite{Lastovicka:2002hw} introduces 1/x and 1 + $Q^2$/$Q_0^2$  as magnification factors. The former confirms that as the probing goes deeper into the proton's structure and x approaches zero, the magnification factor increases accordingly. The latter normalizes the physical dimension of $Q^2$ by dividing it with a constant $Q_0^2$, preventing the issue of $Q^2=0$ by adding 1 to the ratio $Q^2/Q_0^2$. With these preparations in place, all the necessary components are ready to begin depicting the proton's structure function at low x values using the framework of Fractal formalism. The following section is dedicated to outlining the parameterization of quark Parton Distribution Functions (PDFs) within the context of the Fractal framework.

\section{Parametrization of the Structure Function}
\label{parametrization}

The concept of self-similarity, when applied to the structure of the proton's confinement, leads to a simple parametrization of quark densities within the proton. This parametrization is based on Eq. (\ref{twod}), and can be constructed using magnification factors $1/x$ and $1+Q^2/Q_0^2$~\cite{Lastovicka:2002hw}.
The distributions described by Eq. \ref{twod} have linear behavior in log-log space. This linearity is also found for unintegrated sea quark densities   at low $\textit{x}$ and $k_t^2$ \cite{Abdulov:2021ivr}.
An unintegrated quark density may be written in the following general form:

\begin{eqnarray}
\log f_i(x,Q^2)&=&{\cal D}_{1} \cdot \log \frac{1}{x} \cdot \log (1+\frac{Q^2}{Q_0^2}) + {\cal D}_2 \cdot \log \frac{1}{x}\nonumber\\&& + {\cal D}_3 \cdot \log (1+\frac{Q^2}{Q_0^2}) + {\cal D}^i_0
\label{param1}
\end{eqnarray}
where $i$ denotes a quark flavor. Following Eq. (5), ${\cal D}_1$,${\cal D}_2$ and, ${\cal D}_3$ are the  Fractal dimensions related to  $x$ and  $Q^2$ parameters respectively. The normalization parameter ${\cal D}^i_0$ is related to each kind of partons. Conventional, integrated quark densities $q_i(x, Q^2)$ are defined as a sum over all contributions with quark virtualities smaller than that of the photon probe, $Q^2$. Thus, $f_i(x, Q^2)$ has to be integrated over $Q^2$, yielding the following relationship between the integrated and unintegrated quark densities:
\begin{equation}
q_i(x,Q^2) = \int^{Q^2}_0 f_i(x,q^2)\ dq^2 .
\label{param2}
\end{equation}

By Integrating on $f_i(x,q^2)$ in equation (\ref{param2}), one can obtain the following analytical expression for the parametrization of a quark density:
\begin{eqnarray}
q_i(x,Q^2) &=& \frac{{\mathbf e}^{{\cal D}^i_0}~Q_0^2~x^{-{\cal D}_2}}{1+{\cal D}_3-{\cal D}_1\log {x}}\nonumber\\&&\left(x^{-{\cal D}_1\log (1+\frac{Q^2}{Q_0^2})}(1+\frac{Q^2}{Q_0^2})^{{\cal D}_3+1}-1\right) .
\label{param3}
\end{eqnarray}
It is important to note that in this parametrization, only the normalization parameter ${\cal D}_0^i$ depends on the quark flavor, while the other parameters are independent of flavor. This assumption implies that all quarks follow the same Fractal structure, i.e., the dimensions $D_i$ and magnification factors are common for all quarks, and they differ only in normalization.

The proton structure function $F_2$ is directly related to the quark densities, given by $F_2 = x\sum_{i}e_i^2(q_i+\bar q_i)$. Therefore, assuming the flavour symmetry of Eq. (\ref{param3}), we can express $F_2$ directly in terms of the parameters in Eq. (\ref{param3}) by replacing $x^{-{\cal D}_2}$ with $x^{-{\cal D}_2+1}$ and introducing a common normalisation factor $\mathbf{e}^{{\cal D}_0}$, as follows:
\begin{eqnarray}
F_2(x,Q^2) &=& \frac{{\mathbf e}^{{\cal D}_0}~Q_0^2~x^{-{\cal D}_2+1}}{1+{\cal D}_3-{\cal D}_1\log {x}}\nonumber\\&&\left(x^{-{\cal D}_1\log (1+\frac{Q^2}{Q_0^2})}(1+\frac{Q^2}{Q_0^2})^{{\cal D}_3+1}-1\right).\nonumber\\ 
F_L(x,Q^2) &=&F_2(x,Q^2)\times\frac{R}{1+R}
\label{paramf2}
\end{eqnarray}
A free parameter \textit{R} will be set by global analysis. To investigate the internal structure of the proton at low values of $x$, one must analyze the reduced cross-section data obtained from $e^{\pm}p$ Deep Inelastic Scattering (DIS) processes at low $x$. For unpolarized $e^{\pm}p$ scattering, the reduced cross-section at low $Q^2$ values, i.e. $Q^2 \ll M_Z^2$, can be expressed as \cite{H1:2010fzx}:

\begin{equation}\label{Eq:eq1}
\sigma_{r,NC} = F_2(x,Q^2) - \frac{y^2}{Y_+} F_{\rm L}(x,Q^2)~,
\end{equation}
where the kinematic variables $x$, $Q^2$, and $y$ are defined as:

\begin{equation}
Q^2=-q^2,:x=\frac{Q^2}{2(P.q)},:y=\frac{(P.q)}{P.k},
\end{equation}
Here, $P$, $k$, and $q$ denote the four-momentum of the incoming proton, incoming lepton, and exchanged boson, respectively, and $Y_+=1+(1-y)^2$. It should be noted that for $y$ values larger than approximately 0.5, the contribution of the longitudinal structure function $F_{\rm L}$ becomes significant.
The detailed information for each data set is summarized  in Table~\ref{AllData}.  
\begin{table*}
\caption{\label{tab:par} The numerical values and their uncertainties extracted for parameters  related to fit-4serires data, fit-3serires data,  fit-2serires data, and fit-1serires data .}
\begin{center}
	\begin{tabular}{lllll}\hline
		Parameter   & fit-4ser. data & fit-3ser. data & fit-2ser. data & fit-1ser. data  \\ \hline
		'${\cal D}_0$' & $1.60 \pm 0.24$& $1.83 \pm 0.31$& $1.95 \pm 0.36$& $1.69 \pm 0.31$  \\ 
		'${\cal D}_1$' & $0.0807 \pm 0.0016$& $0.0793 \pm 0.0016$& $0.0793 \pm 0.0017$& $0.0798 \pm 0.0019$  \\ 
		'${\cal D}_2$' & $0.992 \pm 0.010$& $0.986 \pm 0.012$& $0.982 \pm 0.013$& $0.991 \pm 0.013$  \\ 
		'${\cal D}_3$' & $-1.339 \pm 0.013$& $-1.321 \pm 0.014$& $-1.318 \pm 0.014$& $-1.330 \pm 0.015$  \\ 
		'$Q_0^2$' & $0.067 \pm 0.010$& $0.0566 \pm 0.0099$& $0.0524 \pm 0.0099$& $0.062 \pm 0.012$  \\ 
		'$R$' & $0.347 \pm 0.020$& $0.366 \pm 0.022$& $0.383 \pm 0.024$& $0.393 \pm 0.027$  \\ 
		Fit status   & converged & converged & converged & converged  \\ \hline
	\end{tabular}
\end{center}
\end{table*}

\begin{table*}
\caption{\label{tab:par2} The numerical values and their uncertainties extracted for parameters  related to  Ref.\cite{Lastovicka:2002hw} and  Ref.\cite{Mohsenabadi:2021vvj} .}
\begin{center}
	\begin{tabular}{lll}\hline
		Parameter   & Ref.\cite{Lastovicka:2002hw} & Ref.\cite{Mohsenabadi:2021vvj}  \\ \hline
		'${\cal D}_0$' & $0.330 \pm 0.195$& $4.81 \pm 0.01$  \\ 
		'${\cal D}_1$' & $0.073 \pm 0.001$& $-0.0051 \pm 0.00009 $ \\ 
		'${\cal D}_2$' & $1.013 \pm 0.01$& $1.138 \pm 0.002 $ \\ 
		'${\cal D}_3$' & $-1.287 \pm 0.01$& $-1.285 \pm 0.005$  \\ 
		'$Q_0^2$' & $0.062 \pm 0.010$& $0.0187 \pm 0.0001$  \\ 
		Fit status   & converged & converged  \\ \hline
	\end{tabular}
\end{center}
\end{table*}

\begin{table*}
\caption{\label{tab:Chi2}The numerical results for the correlated $\chi^2$, log penalty $\chi^2$, total $\chi^2$ and the total $\chi^2$/ degree of freedom (dof)  of each data sets for different fit-4serires data, fit-3serires data,  fit-2serires data and fit-1serires data.}

\begin{center}
	\begin{tabular}{lllll}\hline
		Dataset     & fit-4ser. data   & fit-3ser. data   &fit-2ser. data   & fit-1ser. data  \\ \hline
		HERA1+2 NCep 820 & 67 / 41& 67 / 41& 67 / 41& -   \\ 
		HERA1+2 NCep 460 & 144 / 111& - & - & -   \\ 
		HERA1+2 NCep 575 & 155 / 143& 161 / 143& - & -   \\ 
		HERA1+2 NCep 920 & 241 / 211& 241 / 211& 244 / 211& 240 / 211  \\ 
		Correlated $\chi^2$  & 73& 56& 39& 33  \\ 
		Log penalty $\chi^2$  & +0.69& +1.6& +2.2& +3.2  \\ 
		Total $\chi^2$ / dof  & 680 / 500& 527 / 389& 353 / 246& 276 / 205  \\ \hline
	\end{tabular}
\end{center}
\end{table*}

\section{Fitting Contents and results}
\label{fitting}
In this section, we utilize experimental data obtained from HERA to study electron(positron)-proton scattering in DIS processes within the xFitter framework. The H1 and ZEUS Collaborations investigated a vast kinematical phase space in ($x, Q^2$), where the experimental data covered $0.005<x<0.65$ and $0.045<Q^2<50000GeV^2$ for neutral current (NC) interactions and $0.01<x<0.4$ and $200<Q^2<50000GeV^2$ for charged current (CC) interactions, respectively. Since we focus on the low-$x$ region, we select a subset of the combined HERA data related to NC interactions in the $x<0.01$ region. The total number of experimental data used in our analysis is $N_{data}=500$, and we summarize them in Table~\ref{AllData}.\\ 
The following step involves a brief explanation of   error estimation  and the results we obtained by xFitter \cite{Abdolmaleki:2019tbb,Tooran:2019cfz,Vafaee:2017nze}. xFitter is a powerful tool that provides a flexible platform for fitting various types of parton distribution functions (PDFs) to experimental data. It includes a variety of PDF sets, including those based on different parameterizations and fitting methods, as well as a set of commonly used models for the strong coupling constant and heavy quark masses. Additionally, xFitter supports the implementation of different types of systematic uncertainties in the fit, such as experimental and theoretical uncertainties. To estimate the uncertainties associated with the experimental data used in the fit, xFitter provides various methods, such as the Hessian method and Monte Carlo methods. The Hessian method involves calculating the second derivatives of the $\chi^2$ function with respect to the fit parameters at the minimum $\chi^2$ point, which can be used to estimate the PDF uncertainties. On the other hand, Monte Carlo methods involve generating a large number of pseudo-data sets by adding Gaussian-distributed random errors to the experimental data and then fitting each of them to obtain a set of PDF replicas. The spread of the resulting PDF replicas can be used to estimate the PDF uncertainties.
It is important to carefully evaluate the uncertainties associated with the experimental data to ensure that they are properly taken into account in the fit. This can help to obtain reliable PDFs and associated uncertainties, which are essential for making predictions for a wide range of high-energy physics observable.
The $\chi^2$-function is a tool to determine how well a particular QCD model fits experimental measurements. In our analysis, we aim to obtain the best values of 6 independent free parameters by minimizing the $\chi^2$-function. When all of the correlated uncertainties associated with experimental measurements are known, the $\chi^2$-function in the xFitter framework is given by \cite{Aaron:2009aa}:
\begin{equation}
\chi^2 = \sum_i \frac{ [d_i -  t_i(1-\sum_j \beta_{j}^{i} s_j )]^2 }{\delta^2_{i,unc} t^2_i +\delta^2_{i,stat} d_i t_i } + \sum_j s^2_j\;,
\label{eq:chi2}
\end{equation}
The $\chi^2$-function presented above calculates the goodness of fit between theoretical predictions $t_i$ and experimental measurements $d_i$ for each data point $i$, taking into account statistical and systematic uncertainties. The systematic uncertainties are categorized as correlated and uncorrelated, with $\beta_{j}^{i}$ and $s_j$ representing the corresponding uncertainties and nuisance parameters, respectively. The $\delta^2_{i, stat}$ and $\delta^2_{i, unc}$ terms describe the relative statistical and uncorrelated systematic uncertainties.
In addition to obtaining the central values of the 6 free parameters, it is essential to determine their uncertainties. The uncertainties of parton distribution functions have been estimated in several studies by the Hessian method, such as Refs.~\cite{Hirai:2007cx,Pumplin:2001ct, Pumplin:2002vw,Martin:2002aw,Blumlein:2002qeu, Hirai:2006sr,Leader:2005ci,deFlorian:2005mw, Hirai:2004wq}. One of the most commonly used approaches for estimating uncertainties is the Hessian minimum iteration method. We utilize this method in our QCD analysis and details of this method are available in these references. 
Four different fits with the name fit-4serires data, fit-3serires data, fit-2serires data, and fit-1serires data are introduced  in Table~\ref{tab:par}.
The first data set,fit-1serires data,  only contains HERA I+II-920 data to prepare a fine base for investigating the impact of other data sets on PDFs. In the second data set,fit-2serires data, the HERA I+II-820 data are added to the first data set.  The third data set, fit-3serires data, and the HERA I+II-575 data are added to the second data set. Finally, the third data set,fit-4serires data, and the HERA I+II-460 data are added to contain all previously mentioned data. According to Table~\ref{tab:Chi2} the extracted values of $\chi^2$/dof  for (fit-1serires data) are 1.34 , (fit-1serires data) 1.43 ,(fit-3serires data) 1.35 and (fit-4serires data) 1.36 , respectively.\\
We have also a comparison of our results with those from  \cite{Lastovicka:2002hw} and \cite{Mohsenabadi:2021vvj} in Table~\ref{tab:par2}. The results show an agreement with those from \cite{Lastovicka:2002hw} because we did the same  but in the frame of xFitter and also including new data from HERA2015 \cite{Abramowicz:2015mha}. The results show a negative value for Fractal dimension of $ \mathcal{D}_3$. To see why,  we revisit Equation 6 in this paper and also refer to \cite{Abdulov:2021ivr}. In \cite{Abdulov:2021ivr}, the authors demonstrate that the up quark parton distribution function exhibits a linear behavior with a negative slope in log-log space at the low  x  region and also for  $Q^2$  greater than  1  ${GeV}^2$  (see Figures 1 and 2 in \cite{Abdulov:2021ivr}). This implies that the quark distribution functions in this region can be described by lines with negative slopes. A line in log-log space corresponds to a power-law function in ordinary space. We then parameterized the quark distribution functions in this region in Eq. (6) with three parameters, $ \mathcal{D}_2$,$ \mathcal{D}_3$ and, $ \mathcal{D}_1$  , which represent the 
Fractal dimensions related to  x and $ Q^2$  and, their correlations respectively. Our results also shows this behavior for quark distribution functions at the low  x  region, exhibiting linearity with a negative slope. Note that our PDFs in Eq.  (6) are parameterized as a function of  $\frac{1}{x} $, not  x , and thus we obtain a positive value for  $\mathcal{D}_2$ . These PDFs are parameterized as a function of  $Q^2$  as $ \mathcal{D}_3 \cdot \log(1+\frac{Q^2}{Q_0^2})$ , from which we obtain the negative value for $\mathcal{D}_3$.
\\
In Fig.~\ref{fig:fig1}, the predictions for the reduced cross-section in the low \textit{x} region below $x<0.001$ for $Q^2=3.5 GeV^2$, $Q^2=2 GeV^2$, and  $Q^2=25 GeV^2$ at $E_p=460GeV$ are presented. Data points are from NC interactions in HERA positron-proton DIS processes for fit-4 series data. We also have a comparison between the theoretical model and experimental data.
In Fig.~\ref{fig:fig2} To evaluate the reduced cross-section in the low \textit{x} region, specifically below $x<0.001$, for three different values of $Q^2$ ($Q^2=8.5 GeV^2$, $Q^2=12 GeV^2$, $Q^2=15 GeV^2$ and $Q^2=20 GeV^2$) at a proton energy of $E_p=575GeV$, data from neutral current (NC) interactions in positron-proton DIS processes at HERA are utilized. The experimental data are available as two sets, fit-3series and fit-4series, and are used to compare theoretical models with the measured values.\\
In Fig.~\ref{fig:fig3}, to predict the reduced cross-section in the low \textit{x} region, specifically below $x<0.0001$, at a proton energy of $E_p=820GeV$, data from neutral current (NC) interactions in positron-proton DIS processes at HERA are utilized. The experimental data are available as three sets: fit-2serires data, fit-3serires data, and fit-4serires data.\\
In Fig.~\ref{fig:fig4} To make predictions for the reduced cross-section in the low \textit{x} region, specifically below $x<0.01$, data from neutral current (NC) interactions in positron-proton DIS processes at HERA are utilized. The experimental data points are available as four sets: fit-1serires data, fit-2serires data, fit-3serires data, and fit-4serires data.\\
In Fig~\ref{xq} we show the behavior of the Fractal quark density for fit-4serires data at $Q^2=2 GeV^2$, $Q^2=5 GeV^2$, $Q^2=10 GeV^2$ and $Q^2=100 GeV^2$.  Finally in Fig~\ref{xq2}, we compare our results with those with did at NLO approximation with including the gluon distribution function at low x ~\cite{Mohsenabadi:2021vvj}. \\
This work demonstrates that the Fractal approach is a promising method to investigate the hadron structure at the low \textit{x} region. We used the reduced cross-section at the LO approximation and only considered the contribution of sea quarks at the low \textit{x} region, neglecting the contribution of gluons. We also used the xFitter framework to study the effect of new data on the results.


\section{Summary and Conclusions}
\label{summary}
In this paper, we have presented the Fractal parton distribution functions, including HERA I+II  DIS experimental data as a base data set  to investigate the proton structure at the low \textit{x} region in the \textbf{xFitter} framework. Our analysis incorporates the HERA I+II Deep Inelastic Scattering (DIS) experimental data as a foundational dataset to scrutinize the proton structure within the low  x ($x<0.01$) and $Q^2>1 GeV^2$ domain, employing the xFitter framework. By juxtaposing the proton's reduced cross-section with empirical data, we find that the Fractal approach provides a plausible description of the physics at low  x. This comparison not only bolsters the validity of the Fractal model but also enhances our understanding of the proton's behavior under these conditions. A comparison of the proton's reduced cross-section with experimental data shows that the Fractal approach may describe the low \textit{x} physics well. Investigating PDFs at low x helps in understanding the non-perturbative aspects of QCD and the dynamics of parton interactions in this regime. The behavior of PDFs at low x also influences the parton-parton collision cross sections, which are fundamental for calculating various scattering processes in particle physics. The results indicate that the Fractal approach is a promising method, offering novel insights into the proton structure at low x  regions and exploring new frontiers in physics.

\section*{Acknowledgments}
Fatemeh Taghavi-Shahri  and  S. Shoeibi acknowledge the Ferdowsi University of Mashhad for the provided facilities to do this project. S. Shoeibi and Shahin Atashbar Tehrani  are grateful to the School of Particles and Accelerators, Institute for Research in Fundamental Sciences (IPM) to make the required support to do this project. Authors also appreciate Hamed Abdolmaleki for useful discussion about \textbf{xFitter}.

\end{document}